\documentclass[12pt]{article}

\usepackage{times}
\usepackage{graphicx}

\newenvironment{sciabstract}{%
\begin{quote} \bf}
{\end{quote}}

\newcounter{lastnote}

\title{Fragmentation in Massive Star Formation}

\author
{Henrik Beuther,$^{1\ast}$ Peter Schilke,$^{2}$\\
\\
\normalsize{$^{1}$Harvard-Smithsonian Center for Astrophysics, 60 Garden Street, Cambridge MA02138, USA}\\
\normalsize{$^{2}$Max-Planck-Institute for Astrophysics, Auf dem Huegel 69, 53121 Bonn, Germany}\\
\\
\normalsize{$^\ast$E-mail: hbeuther@cfa.harvard.edu.}
}

\date{}

\begin{document} 


\baselineskip24pt


\maketitle


\begin{sciabstract}
Studies of evolved massive stars indicate that they form in a
clustered mode.  During the earliest evolutionary stages, these
regions are embedded within their natal cores. Here, we show
high-spatial-resolution interferometric dust continuum observations
disentangling the cluster-like structure of a young massive
star-forming region. The derived protocluster mass distribution is
consistent with the stellar initial mass function. Thus, fragmentation
of the initial massive cores may determine the initial mass function
and the masses of the final stars. This implies that stars of all
masses can form via accretion processes, and coalescence of
intermediate-mass protostars appears not to be necessary.
\end{sciabstract}

There is a general consensus that massive stars ($> 8$\,M$_{\odot}$;
solar masses) form exclusively in a clustered mode but the detailed
physical processes are far from clear. While high enough accretion
rates and accretion through disks are capable of forming massive
stars, scenarios like the merging of intermediate-mass protostars at
the dense centers of evolving clusters are also possible
(1-3). Furthermore, one needs to understand why the stellar clusters
have a universal mass spectrum fairly independent of environmental
conditions, and how this mass spectrum evolves. Therefore, it is
crucial to study the earliest evolutionary stages at high spatial
resolution, preferably in the mm-wavelength regime where dust emission
is strong and optically thin, tracing all dust along the line of
sight. The dust emission is directly proportional to the column
density of dense gas within the regions, thus observing the mm
continuum emission in very young massive star-forming regions allows
us to study the gas and dust distributions, possible fragmentation of
the larger scale cores, and physical parameters like masses and column
densities.

Recently, we imaged dust continuum emission at 1.3\,mm and 3\,mm from
the massive star-forming region IRAS\,19410+2336 with the Plateau de
Bure Interferometer [PdBI, (4)].  The region
IRAS\,19410+2336 is in an early stage of high-mass star formation
prior to forming a hot core~-- a dense, hot clump of gas heated by a
massive protostar (5). It is at a distance of $\sim 2$
kiloparsec (kpc) and has an integrated bolometric luminosity of about
$10^4$\,L$_{\odot}$ (solar luminosities). The region is part of a
large sample of high-mass protostellar objects that has been studied
extensively from cm to x-ray wavelengths (6-10).

The PdBI consists of six 15\,m antennas, and we have observed the
source in three different configurations with projected baseline
lengths between 15\,m and 330\,m. The two-dimensional representation
of projected baselines on the plane of the sky~-- the uv-plane~-- is
covered extremely well in this range providing a high image fidelity
at the corresponding spatial frequencies (11). At
1.3\,mm the synthesized beam is $1.5''\times 1''$ and at 3\,mm
$5.5''\times 3.5''$. Additionally, we present single-dish 1.2\,mm
observations of the same region obtained with the IRAM 30\,m telescope
at $11''$ spatial resolution (7). Thus, we are able to
analyze the evolving cluster at several spatial scales down to a
linear resolution of 2000 astronomical units (AU, $1''$ at a distance
of 2 kpc).

The large scale emission observed at a wavelength of 1.2\,mm with the
IRAM 30\,m telescope (Fig. 1a) shows two massive gas cores roughly
aligned in a north-south direction. Based on the single-dish intensity
profiles, we predicted that the cores should split up into
sub-structures at scales between $3''$ and $5''$
(7). PdBI 3\,mm data at more than twice the spatial
resolution show that both sources split up into sub-structures at the
previously predicted scales, about four sources in the southern core
and four in the northern core (Fig, 1b).  At the highest spatial
resolution (Figs. 1c and d), we observe that previously known gas
clumps resolve into even more sub-sources. We find small clusters of
gas and dust condensations with at least 12 sources per large scale
core. Each of the protoclusters is dominated by one central massive
source and surrounded by a cluster of less massive sources. This
provides evidence for the fragmentation of a high-mass protocluster
down to scales of 2000\,AU at the earliest evolutionary stages.

In addition to a morphological interpretation, the data allow a
quantitative analysis of the mass distribution. Assuming that the mm
continuum is produced by optically thin thermal dust emission, one can
calculate the masses and the column densities of the sub-sources
following (12). The evolving protocluster is in a very early
evolutionary state, and only the strongest source in the south
exhibits weak (1\,mJy) cm emission indicative of a recently ignited
star (6,9). This cm emission is unresolved and does not affect the
morphology of the dust emission.  There could be temperature
differences between various clumps as well as temperature gradients
within individual clumps (13). However, in a massive cluster at the
given distance it is difficult to obtain with current observational
capabilities individual temperature estimates for each sub-source, let
alone to derive internal clump temperature gradients. Nevertheless,
due to the early evolutionary stage of the region prior to forming a
significant hot core the dust temperatures throughout the cluster
should not vary too strongly, and it is plausible to assume the same
dust temperature for all sub-sources. Based on IRAS far-infrared
observations, we estimate the average dust temperature to be around
46\,K (6), the dust opacity index beta is set to 2 (7). The
single-dish data (Figure 1a) reveal the overall gas mass, we derive
840\,M$_{\odot}$ in the south and 190\,M$_{\odot}$ in the
north. Calculating the total masses for the southern and northern
cluster from the interferometric data, we get lower values because the
interferometers filter out the large scale emission and trace only the
most compact sources (11). This effect increases with
decreasing wavelength. Thus, the 3\,mm observations (Fig. 1b) still
detect 210\,M$_{\odot}$ in the south and 80\,M$_{\odot}$ in the north,
whereas at 1.3\,mm we only observe the most compact condensations with
a total mass of 98\,M$_{\odot}$ in the south and 42\,M$_{\odot}$ in
the north (Figs. 1c and d). The data show that massive protocluster
evolve in a core-halo fashion where the massive, dense gas clumps from
which the stars are forming are embedded within a larger scale halo of
more broadly distributed gas. A fraction of 80-90$\%$ of the total gas
mass appears to be associated mostly with the halo.

More interesting than the total core masses are the individual masses
and column densities of each sub-source. In the 1.3\,mm PdBI data we
identify 12 individual clumps above the $3\sigma$ level of 9\,mJy/beam
in each of the southern and northern cores. At the assumed temperature
of 46\,K, the $3\sigma$ level corresponds to a clump-mass sensitivity
of $\sim 1.7$\,M$_{\odot}$, and the derived clump masses range between
1.7\,M$_{\odot}$ and 25\,M$_{\odot}$. The calculated peak column
densities are of the order $10^{24}$cm$^{-2}$ corresponding to a
visual extinction $A_{\rm{v}}$ of about 1000. Such an extinction is
too high to be penetrated by near-infrared, mid-infrared or even hard
x-ray emission. Our sample is sensitivity limited for for individual
gas clumps below 1.7\,M$_{\odot}$.  We do not believe that the spatial
filtering property of the interfermetric observing technique affects
our results, because the scales of all, even the most massive clumps
(of order a few arcsec) is far smaller than the spatial structures
filtered out (of sizes above 20 arcsec).  Consequently, only a
large-scale halo common to all sources is affected by the filtering,
while the sources we are interested in are not.

Combining the data from both clusters, we derive a mass spectrum of
the protocluster (Figure 2).  The best fit to the data results in a
mass spectrum $\Delta N/\Delta M \sim M^{-a}$ with a power-law index
$a=2.5$ and a mean deviation $da=0.3$. A potential uncertainty
for the slope of the spectrum is the assumption of uniform dust
temperatures for all sub-sources: while higher temperatures for the
more massive clumps would decrease these derived masses, lower
temperatures for the less massive clumps would increase those
mass estimates. These effects would result in a somewhat flattened
slope. However, as argued before, we infer from the early evolutionary
state of IRAS\,19410+2336 that the dust temperature distribution
should not vary strongly, and we conclude that the relative
accuracy between the derived clump masses~-- and thus the slope of the
mass spectrum~-- is high. This derived power-law index $a$ can directly
be compared with the initial mass function (IMF) results obtained for
more evolved clusters. While there has been a lot of discussion about
the very low-mass end ($<1$\,M$_{\odot}$) there is a general consensus
that the IMF for stars $>1$\,M$_{\odot}$ can be approximated by
$a=2.5\pm 0.2$ (1,14-17). Furthermore,
in a study of low-mass protostars in $\rho$ Ophiuchus, it was found
that the mass spectrum of protostars between 0.5\,M$_{\odot}$ and
3\,M$_{\odot}$ can be approximated by $a=2.5$ (18). While
the observations of $\rho$ Ophiuchus already suggested that the IMF of
low-mass stars is determined at early evolutionary stages, this was
far from clear for massive clusters because merging of
intermediate-mass protostars could form the IMF at later stages as
well (2). Our new data now indicate that the IMF of
high-mass clusters is also determined at the earliest stages of
evolution.  An IMF determined at the very beginning of the cluster
formation is support for the disk-accretion scenario for stars of all
masses, because coalescence would probably occur at later stages and
could hardly know anything about the initial fragmentation
processes. However, these data do not rule out the possibility that
coalescence of intermediate-mass protostars might occur within the
dense centers of individual sub-cores.

{\bf \Large References and Notes}\\
1. R.E. Pudritz, Science 295, 68 (2002).\\
2. S.W. Stahler, F. Palla, P.T.P. Ho. Protostars and Planets IV pp.327-+ (2000).\\
3. C.F. McKee, J.C. Tan. Nature 416. 59 (2002).\\
4. S. Guilloteau, et al., A\&A 262, 624 (1992).\\
5. S. Kurtz, R. Cesaroni, E. Churchwell, P. Hofner, C.M. Walmsley, Protostars and Planets IV pp. 299-+ (2000).\\
6. T.K. Sridharan, H. Beuther, P. Schilke, K.M. Menten, F. Wyrowski, ApJ 566, 931 (2002).\\
7. H. Beuther, et al., ApJ 566, 945 (2002).\\
8. H. Beuther, et al., A\&A 383, 892 (2002).\\
9. H. Beuther, J. Kerp, T. Preibisch, T. Stanke, P. Schilke, A\&A 395, 169 (2002).\\
10. H. Beuther, P. Schilke, T. Stanke, A\&A 408, 601 (2003).\\
11. G.B. Taylor, C.L. Carilli, R.A. Perley, eds., Synthesis Imaging in Radio Astronomy II (1999).\\
12. R.H. Hildebrand, QJRAS 24, 267 (1983).\\
13. F.C. Adams, F.H. Shu, ApJ 296, 655 (1985).\\
14. E.E. Salpeter, ApJ 121, 161 (1955).\\
15. G.E. Miller, J.M. Scalo, ApJS 41, 513 (1979).\\
16. J. Scalo, ASP Conf. Ser. 142: The Stellar Initial Mass Function (38th Herstmonceux Conference) (1998), pp. 201-+.\\
17. P. Kroupa, C.A. Tout, Gilmore G., MNRAS 262, 545 (1993).\\
18. F. Motte, P. Andre, R. Neri, A\&A 336, 150 (1998)\\
19. Based on observations with the IRAM Plateau de Bure Interferometer and
the 30\,m. IRAM is supported by INSU/CNRS (France), MPG (Germany), and
IGN (Spain). We thank an unknown referee for very helpful comments
improving the paper. H.B. acknowledges financial support by the
Emmy-Noether-Programm of the Deutsche Forschungsgemeinschaft (DFG,
grant BE2578/1).\\

\includegraphics[angle=-90,width=16cm]{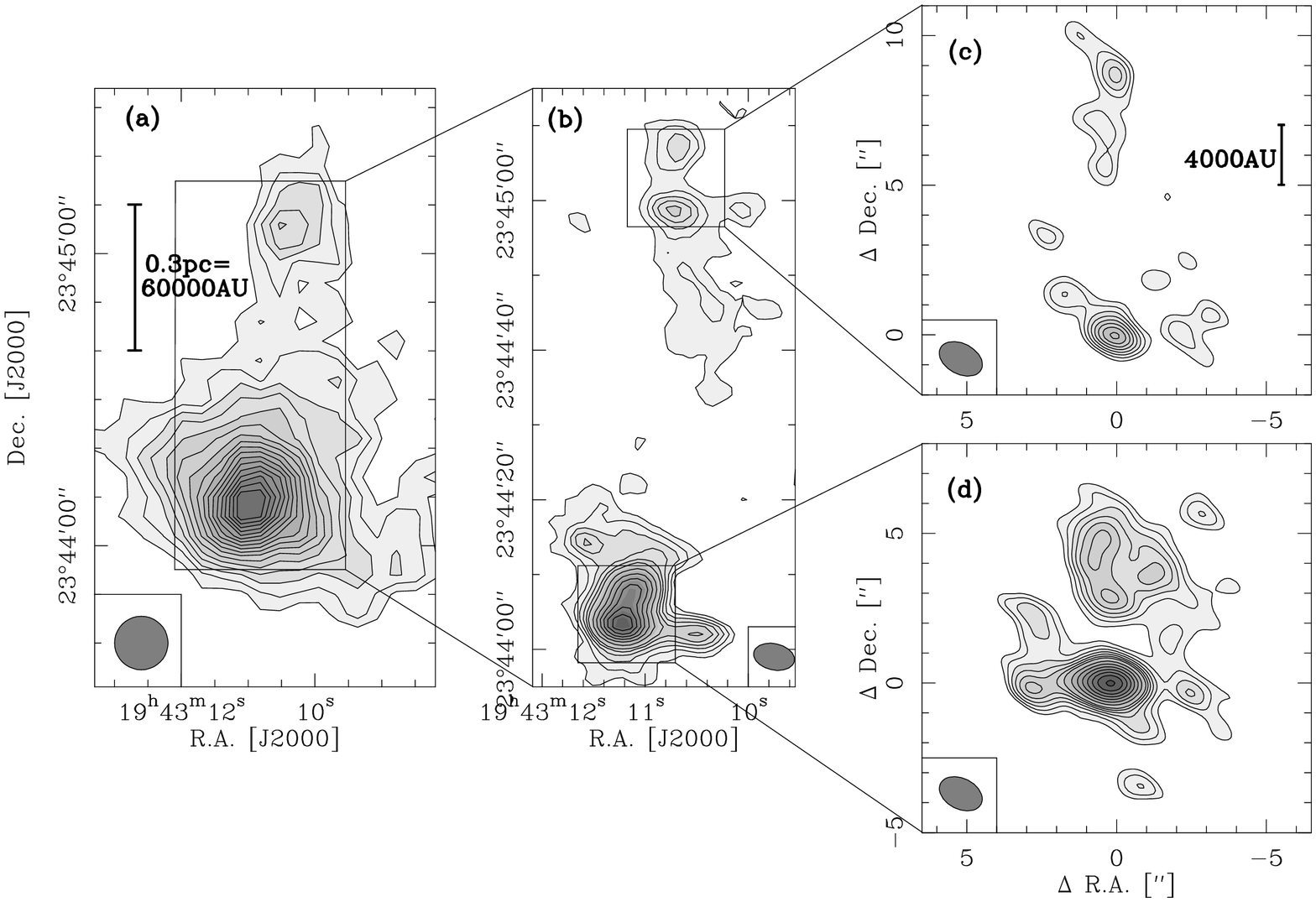}
Fig. 1: Dust continuum images of IRAS\,19410+2336. The left image 
shows 1.2\,mm single-dish data obtained with the IRAM 30\,m at low
spatial resolution (7). The middle and right images
present the 3\,mm and 1.3\,mm PdBI data obtained with a spatial
resolution nearly an order of magnitude better. The beams are shown at
the bottom left or right of each panel, respectively. The contouring
of (a) starts at $15\%$ of the peak flux increasing in $5\%$
intervals, (b) is contoured in $5\%$ intervals of the peak flux
between $5\%$ and $25\%$, and in $10\%$ intervals between $30\%$ and
$90\%$. The 1.3\,mm images in Figures (c) and (d) are contoured at
$1\sigma$ steps between the $3\sigma$ level of 9\,mJy and 27\,mJy, and
at $2\sigma$ steps above that.\\

\includegraphics[angle=-90,width=16cm]{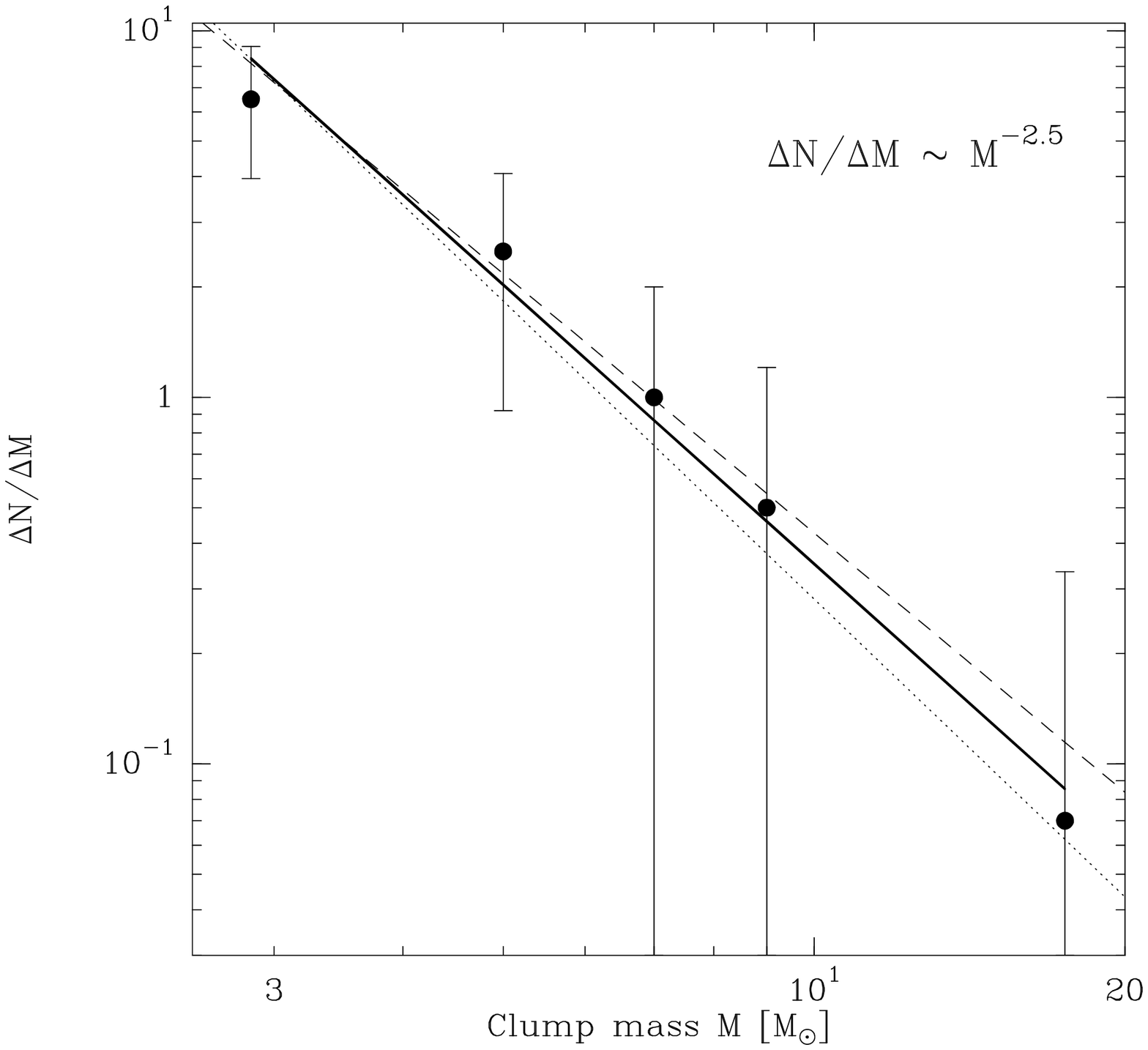} Fig. 2: The mass
Fig. 2: The mass spectrum of IRAS\,19410+2336. The clump-mass bins are
[1.7($3\sigma$),4], [4,6], [6,8], [8,10] and [10,25] \,M$_{\odot}$,
the axes are in logarithmic units. The error-bars represent the
standard deviation of a Poisson-distribution $\sqrt{\Delta N / \Delta
M}$.  The solid line shows the best fit to the data $\Delta N/\Delta M
\sim M^{-a}$ with $a=2.5$. The dashed and dotted lines present the
IMFs derived from Salpeter with $a=2.35$ (14) and Scalo
with $a=2.7$ (16), respectively.

\end{document}